\documentclass[a4paper,11pt]{article}
\pdfoutput=1 
\usepackage{tikz-cd}
\usepackage{jheppub} 
                     
\usepackage{caption}
\usepackage{subcaption}
\usepackage{graphicx}
\usepackage{braket}
\usepackage{xcolor}
\usepackage[T1]{fontenc}
\usepackage{scalerel}
\mdseries 
\usepackage{amsmath}
\makeatother
\usepackage{mathtools}
\newcommand\fat[1]{\ThisStyle{\ooalign{%
  \kern.46pt$\SavedStyle#1$\cr\kern.33pt$\SavedStyle#1$\cr%
  \kern.2pt$\SavedStyle#1$\cr$\SavedStyle#1$}}}

\usepackage{setspace}
\usepackage{xcolor}
\usepackage{hyperref}
\usepackage{mathtools}
\usepackage{esint}
\usepackage{amssymb}
\usepackage{eucal}
\usepackage{graphicx}
\usepackage{amsmath}
\usepackage{mathtools,slashed}

\usepackage{amsmath}

\makeatletter
\newenvironment{sqcases}{%
  \matrix@check\sqcases\env@sqcases
}{%
  \endarray\right.%
}
\def\env@sqcases{%
  \let\@ifnextchar\new@ifnextchar
  \left\lbrack
  \def\arraystretch{1.2}%
  \array{@{}l@{\quad}l@{}}%
}
\makeatother

\usepackage{amsmath}

\usepackage{pstricks-add}
\usepackage[T1]{fontenc} 
\usepackage{amsfonts}
\usepackage{tikz}
\usetikzlibrary{positioning}
\title{{\textbf{$\fat{T\bar{T}}$-deformed 2D Yang-Mills at large N:\\
collective field theory and phase transitions}}}

\newcommand{\RNumb}[1]{\uppercase\expandafter{\romannumeral #1\relax}}
\author[a,b]{A. Gorsky,}
\author[b]{D. Pavshinkin,}
\author[b,c]{and A. Tyutyakina}

\affiliation[a]{Institute for Information Transmission Problems RAS, 127051 Moscow, Russia}
\affiliation[b]{Moscow Institute of Physics and Technology, Dolgoprudny 141700, Russia}
\affiliation[c]{Russian Quantum Center, Skolkovo, Moscow 143025, Russia}

\abstract{We consider the $T\bar T$ deformation of 2d large $N$ YM theory on a cylinder, sphere and disk. The collective field theory
Hamiltonian for the deformed theory is derived and the particular solutions 
to the equations of motion of the collective theory  are found for the sphere. The account of the non-perturbative 
branch of the solution amounts to the first-order phase transition at the $(A,\tau)$ plane. 
We analyze the third-order phase transition in the deformed theory on the disk  and derive the critical area
as a function of the boundary holonomy. A kind of Hagedorn behavior in the spectral density is discussed.}

\emailAdd{shuragor@mail.ru, dmitriy.pavshinkin@phystech.edu, tyutyakina.av@phystech.edu}

\begin{document} 
\maketitle
\flushbottom

\section{Introduction}

The integrable $T\bar T$ deformation of 2d field theories \cite{Zam} attracts a lot of attention as an exceptional example of an irrelevant operator acting in a controlled way in the UV. In particular, energy levels of a theory on a cylinder are governed by the well studied Burgers' differential equation and the exact results concerning an S-matrix are available \cite{SmirZam,Tateo,DubGorbMirb}. The closed form of $T\bar T$ deformed theories in terms of solutions of Burgers' equation in many particular specific models was explored in \cite{closed}.
Within the framework of the holographic description, turning on the $T\bar T$ with one sign of the deformation parameter $\tau$ moves the boundary CFT$_2$ inside the bulk, and the spectrum of a resulting QFT appears to be truncated from above \cite{Moving}.
With another sign, the perturbed theory becomes non-local and  the Hagedorn growth of the density of states takes place \cite{LST}. The deformed quantum mechanics  has been formulated \cite{gross1,gross2} for generic potential and the spectrum of the deformed
theory has been  derived. More recently the attractive qualitative picture behind the  $T\bar T$ deformation has been
developed \cite{cardy,jiang}. It was argued that point-like particles became the finite rods under the deformation
with the positive parameter and the effective length of the system gets decreased. On the other hand for the 
opposite deformation sign the additional space gets opened and the effective size of a system gets increased.

The 2d YM theory at large $N$ is an example of solvable  perturbed topological BF theory \cite{migdal} and admits a kind
of stringy representation \cite{GrossTaylor}. It enjoys the 3-rd order Douglas-Kazakov phase transition at a critical value of the area
for $S^2$ geometry \cite{DK} and a critical relation between the area and the boundary holonomies for the cylinder 
and the disk \cite{GroMat}. Microscopically the transition takes place when instanton solutions
start to dominate the partition function and the topological susceptibility of the ground state becomes finite.
The limit of the large $N$ can be described in terms of the collective field theory with the Das-Jevicki Hamiltonian \cite{das-jevicki-sakita} and
the phase transition can be described in terms of the particular solutions  to the hydrodynamical equations
in the collective field description \cite{GroMat}. 

The 2d YM admits the fermionic description (see, for instance, \cite{MinPol}) which for the case 
of the cylinder can be considered as the result of application of T-duality for $S^1$, the eigenvalues
of holonomies of the gauge field become the coordinates of fermions on the dual circle and the 
components of the electric field become the momenta. The fermionic language provides a new
interpretation of the 3-rd order phase transition in terms of reunion probabilities of the vicious
random walkers at the circle \cite{deharo,fms1,fms2,gmn}. The problem of evaluation of a partition function
gets mapped into the evaluation of  probability of the random process for the vicious walkers on the circle
with the different initial and finite configurations at  fixed time $T$. The strong coupling
phase in this language corresponds to the process when the trajectories of the vicious walkers 
with the windings dominate. The phase transitions in q-YM has been analyzed in \cite{jafferis,
qym,marino05,okuyama}.

Another attractive feature of the fermionic language is the relation of the 2d YM and q-YM 
partition functions with the partition function of the black holes with the particular chemical potentials
for the charges \cite{vafa,osv,ooguri,baby}. It is fermionic language which is the effective tool in separation
of the 2d YM partition function into the left and right sectors suitable for the mapping into BH
and topological strings. 

If the Wilson line in the single-raw representation of $SU(N)$ 
in time direction is included the system of non-interacting fermions gets generalized to the 
$N$-body trigonometric Calogero-Moser-Sutherland system \cite{gn1}. The 2d YM+Wilson line system can be lifted to the
perturbed G/G WZW theory and the Calogero-Moser system is generalized to the Ruijsenaars-Schneider
relativistic version \cite{gn2}. The collective field theory for the  trigonometric Calogero-Moser-Sutherland
system has been developed in \cite{abanov} and has been identified as the bidirectional
Benjamin-Ono equation.

$T\bar T$ deformation of the 2d YM theory has been worked out at the classical level in some different ways, and the deformed Lagrangian has been shown to be a hypergeometric function \cite{BornInf,Shyam,DBI}. In fact the deformation 
has the $\tau T^2$ form.
It looks quite simple in the Hamiltonian formalism and comes down to the renormalization of the quadratic Casimir 
appearing in the heat kernel partition function. As a result, the DK-like phase transition has been found in $T\bar T$-YM on $S^2$ \cite{tierz1}. The deformation preserves the order of the transition and its instanton interpretation but the critical area gets renormalized in a non-trivial way $A\rightarrow Ab_{\infty}(\tau,A)$. 

The DK phase transition for the $T^2$ deformed 2d q-YM theory
has been studied as well \cite{tierz2}. It was emphasized that the free fermion picture 
for the $T^2$ deformed theory is lost. Moreover there is no naive factorization property for the fermions and
the direct mapping to the BH partition function is absent. The reason for the strong deformation of the  fermionic
picture  is due to the nontrivial T-duality transformation in the deformed theory 
which brings into the game the complicated non-local many-body interactions.

In this paper we examine $T\bar T$ deformation of 2d $SU(N)$ YM at large $N$ by turning to the collective field description through a density of the holonomy eigenstates on a circle. The density evolves along the cylinder according to the particular generalization of the complex Hopf equation due to deformation of the Das-Jevicki Hamiltonian.  The exact solution to this equation is found for $S^2$. Thereby, the scaling parameter $b_{\infty}(\tau,A)$ from \cite{tierz1}  is reproduced non-perturbatively in $\tau$. However we show that for $\tau<0$ there are two solutions coexisting on an equal footing 
 and account of both solutions in the partition function indicates a first order phase transition which proceeds at some critical point at the $(A,\tau)$ plane. Note that the necessity to take into account both solutions has been discussed in a slightly different 
context in \cite{verlinde}. 
One more point elaborated in our study concerns the derivation of the critical point for the DK phase transition
in $T^2$ deformed YM theory on a disk. We shall find the critical area for the disk geometry as a
function of a boundary holonomy. We also investigate the phase structure for the q-deformed $T\bar T$-YM on $S^2$ at $p=1$.
 
 The paper is organized as follows.
We start by reviewing the partition function of the $T\bar{T}$-deformed YM on a cylinder in section \ref{sec:2}.
The collective field theory description of the theory on the cylinder is build in section \ref{sec:3.1}. In section \ref{sec:3.2} we solve the resulting hydrodynamical equations for the case of trivial boundary Wilson loops. Then we discuss 3-rd order DK phase transition in the deformed theory on the sphere and disk. The equation for the critical area of the disk with general boundary holonomy is obtained in \ref{sec:4.1}.  In section \ref{sec:5} we find Hagedorn growth of the density of states in $T\bar{T}$-YM on $S^2$.
The new 1-st order phase transition is studied in \ref{sec:6}.  
Section \ref{sec:7} contains the results of numerical calculations of the critical area for $T\bar{T}$-perturbed $q$-YM on $S^2$
at $p=1$.
In section \ref{sec:8} we discuss the results obtained and summarize further research.
Some details concerning the derivation  of the collective theory can be found  in the appendix \ref{Appendix1}. 

\section{$T\bar{T}$-YM on a cylinder}\label{sec:2}

We start this section with a brief overview of the derivation of ${T\bar{T}}$-YM partition function on the cylinder. The Lagrangian density of the pure YM theory for $U(N)$ gauge group on a unit volume surface is as follows
\begin{equation}
\CMcal{L}_0=\frac{1}{2g_{YM}^2} TrF^2
\end{equation}
  
\begin{equation}
\text{in the gauge}~A_0=0:~~~ \Pi^{1}=\frac{\partial \CMcal{L}_0}{\partial (\partial_0 A_{1})}=-\frac{1}{g_{YM}^2}(\partial_0 A_{1}) ,~~~~\Pi^{0}=0
\end{equation}
while the Hamiltonian density is 
\begin{equation}
\CMcal{H}_0=\frac{1}{2g_{YM}^2}(\partial_0 A_{1})^2=-\frac{g_{YM}^2}{2}tr(\Pi^{1})^2.
\end{equation}
Let us place the theory on a cylinder with $ L$-periodic space. The wave function that solves the Gauss law constraint is a class function and depends only on $U=P\exp{\int_0^{ L}A_1dx}$. Thus the Hamiltonian is diagonal in the bases of irreducible characters  \cite{Cohomology}
\begin{equation}
\hat{{H}}_0\chi_R(U)=\frac{Lg_{YM}^2}{2}C_2(R)\chi_R(U), ~~~\hat{{H}}_0=\frac{g_{YM}^2}{2}\int_0^{ L 
}\frac{\delta^2}{(\delta A_1^a)^2}dx.
\end{equation}
$T\bar T$ flow is governed by the following equation in Euclidean signature 
\begin{equation}
\partial_{\tau}\CMcal{L}(\tau)=-detT_{\mu\nu}(\tau),~~~T\bar T(\tau)=-\pi^2detT_{\mu\nu}(\tau)
\end{equation}
where $T_{\mu\nu}(\tau)$ is a stress-energy tensor of the deformed theory. In the case of the 2d YM theory $detT_{\mu\nu}=-T_{00}^2=-\CMcal{H}^2$ so one immediately obtains the deformed Hamiltonian \cite{BornInf}
\begin{equation}\label{eq2.6}
\CMcal{H}_{\tau}=\frac{\CMcal{H}_{0}}{1-\tau\CMcal{H}_{0}}.
\end{equation}

 Following  \cite{migdal} the partition function of the 2d $T\bar{T}$-YM on a cylinder of area $A$ and with the boundary holonomies $U_1$ and $U_2$ 
 can be represented as the sum over irreducible representations of the gauge group
\begin{equation}
Z_N(U_1,U_2|A,\tau)=\sum_R \chi_R(U_1)\chi_R(U_2^{\dagger})\exp\bigg(-\frac{\frac{Ag_{YM}^2}{2}C_2(R)}{1-\tau\frac{g_{YM}^2}{2}C_2(R)}\bigg).
\end{equation} 
The appearance of the combination $Ag_{YM}^2$ is a consequence of the diffeomorphism invariance of 2d YM. The deformed partition function satisfies the following equation 
\begin{equation}
-\partial_{\tau} Z_N=A\partial_A^2Z_N
\end{equation} 
that can be used to derive the expression for the deformed free energy from the original one. Hereafter we set $Ng_{YM}^2=1$ and rescale the deformation parameter $\tau\rightarrow 2\tau/N^2$.

$U(N)$ irreducible representations are labeled by distributions of integers 
\begin{equation}\label{eql}
+\infty>l_1>l_2>...>l_N>-\infty.
\end{equation}
The corresponding Casimir element and character of a unitary matrix $U_1$ are as follows
\begin{equation}\label{eq3}
C_2(R)=\sum_{i=1}^{N}\Big(l_i-\frac{N-1}{2}\Big)^2-\frac{N}{12}(N^2-1),~~~\chi_R(U_1)=\frac{\det\limits_{_{ab}}
\begin{Vmatrix}
  e^{i l_a \theta_b}
\end{Vmatrix}}{\prod\limits_{s<r}(e^{i\theta_s}-e^{i\theta_r})}
\end{equation}
where $e^{i\theta_k}$, $k=1,...N$  denote eigenvalues of $U_1$.
By introducing the new variables $y_k=l_k+\frac{N-1}{2}$, we obtain 

\begin{equation}
Z_N(U_1,U_2|A,\tau)=\frac{1}{N!}\frac{1}{2^{N(N-1)}}\sum_{\{y_k\}}\frac{\det\limits_{_{ab}}\begin{Vmatrix}
  e^{i y_a \theta^{}_b}
\end{Vmatrix}\det\limits_{_{cd}}\begin{Vmatrix}
  e^{-i  y_c \phi^{}_d}
\end{Vmatrix}}{\prod\limits_{s<r}\big[\sin\frac{\theta^{}_s-\theta^{}_r}{2}\sin\frac{\phi^{}_s-\phi^{}_r}{2}\big]}\times
\end{equation}
\begin{equation}\label{eq.231}
\times \exp\bigg({-\frac{A}{2N}\sum_{j=0}^\infty \tau^j \Big[\sum_{k=1}^N y_k^2-\frac{N^2(N-1)}{12}\Big]^{j+1}}\bigg)
\end{equation}
where the sum runs over all unconstrained  $y_k$. 

Since the coordinates  $\text{diag}\{e^{i\theta_1},...e^{i\theta_N}\}$ on the maximal torus of $U(N)$ completely define the class functions $\chi_R(U)$, one can recast the undeformed Hamiltonian as follows \cite{MinPol,douglas}
\begin{equation}
{H}_{0}=\frac{L}{2N\tilde{\Delta}(\theta)}\bigg[-\sum_{k=1}^N \frac{\partial^2}{\partial \theta_k^2}-\frac{N^2(N-1)}{12}\bigg]\tilde{\Delta}(\theta),~~~\tilde{\Delta}(\theta)=\prod\limits_{s<r}\sin\frac{\theta_s-\theta_r}{2}.
\end{equation} 
In this form ${H}_{0}$, obviously describes a system of free fermions on a circle $\theta_i\in[0,2\pi)$, and $y_k$ label their discrete momenta. Fermionic picture provides a straight way to the stringy description of 2d YM \cite{GrossTaylor}.

\section{Collective field theory}\label{3}
In the limit of the large $ N $, the theory is greatly simplified. At least for the case of symmetric boundary conditions $U_1=U_2^{\dagger}$, the partition function is dominated by a specific representation. In turn, this representation determines the saddle distribution of the fermions, which can be described in the weak coupling phase like a one-dimensional fluid on the cylinder. The evolution of this fluid from the initial configuration $\{\theta_i\}$ to the final $\{\phi_i\}$ is governed by the classical Das-Jevicki Hamiltonian. We derive its deformed version in this chapter.
\subsection{Deformation of Das-Jevicki Hamiltonian}\label{sec:3.1}

We obtain the large $N$ asymptotic of $Z_N(U_1,U_2|A,\tau)$  and describe how the boundary Wilson loop evolves along the cylinder \cite{GroMat}.

To begin with, one can make sure that the deformed partition function \eqref{eq.231} satisfy the following modification of the heat equation on the group

\begin{equation}\label{eq6501}
2\frac{\partial Z_N}{\partial A}=-\frac{N^2}{\tilde{\Delta}(\theta)}\sum_{i=0}^\infty \tau^i \bigg(-\frac{1}{N^3}\sum_{k=1}^N \frac{\partial^2}{\partial \theta_k^2}-\frac{1}{12}\bigg)^{i+1}\big[\tilde{\Delta}(\theta^{}) Z_N\big].
\end{equation}
Instead of eigenvalues $e^{i\theta^{}_k}$ consider their distribution $\sigma_{}(\theta)$ that serve as densities of the one-dimensional fluid on the circle of unit radius. We show in appendix \ref{Appendix1} that in order to respect the equation \eqref{eq6501} the leading large $N$ asymptotic of the partition function $ Z=e^{N^2{F}}$ should take the form 
\begin{equation}\nonumber
{F}[\sigma_1(\theta),\sigma_2(\phi)|A,\tau]=-S[\sigma_1(\theta),\sigma_2(\phi)|A,\tau]
\end{equation}
\begin{equation}\label{eqF}
-\frac{1}{2}\int\sigma_1(\theta)\sigma_1(\theta^\prime)\log\Big|\sin\frac{\theta-\theta^\prime}{2}\Big|d\theta d\theta^\prime-\frac{1}{2}\int\sigma_2(\phi)\sigma_2(\phi^\prime)\log\Big|
\sin\frac{\phi-\phi^\prime}{2}\Big|d\phi d\phi^\prime
\end{equation}
where the functional $S$ yields the Hamiltonian evolution of the fluid along the time $A$
\begin{equation}\label{eq15}
\frac{\partial S}{\partial A}+\widetilde{H}=0
\end{equation}
with the Hamiltonian
\begin{equation}\label{eqHam}
\widetilde{ H}
=\frac{H+\frac{1}{24}}{1+2\tau(H+\frac{1}{24})}
\end{equation}
which is $(-2\tau)T^2$ deformation of the Das-Jevicki Hamiltonian for the collective coordinate $\sigma_1(\theta)$
\begin{equation}
H[\sigma_1(\theta),\Pi(\theta)]=\frac{1}{2}\int_0^{2\pi}\sigma_1(\theta)\bigg[\Big(\frac{\partial \Pi}{\partial\theta}\Big)^2-\frac{\pi^2}{3}\sigma_1^2(\theta)\bigg]d\theta,~~~\Pi(\theta)=\frac{\delta S}{\delta\sigma_1(\theta)}.
\end{equation} 
$\Pi(\theta)$ and $\sigma_1(\theta)$ are canonically conjugate variables. The details of the derivation of 
deformed Das-Jevicki Hamiltonian are presented in the appendix \ref{Appendix1}. The Hamiltonian equations 
\begin{equation}
\begin{cases}
\frac{\partial \sigma(\theta)}{\partial t }=\frac{\delta \widetilde{H}(\sigma , \Pi) }{\delta \Pi (\theta) } =\alpha \frac{\delta H(\sigma , \Pi) }{\delta \Pi (\theta) },\\
\frac{\partial \Pi(\theta)}{\partial t }=-\frac{\delta \widetilde{H}(\sigma , \Pi) }{\delta \sigma (\theta) }=-\alpha\frac{\delta H(\sigma , \Pi) }{\delta \sigma (\theta) }
\end{cases}~~~
\alpha=\frac{\partial \widetilde{ H}}{\partial H}=\frac{1}{\big[1+2\tau(  H+\frac{1}{24})\big]^2}
\end{equation}
amount to the system of Euler's equations with the corresponding boundary conditions
\begin{equation}\label{eq17}
\begin{cases}
\frac{\partial \sigma}{\partial t}+\alpha \frac{\partial (\sigma v)}{\partial \theta}=0 \\
\frac{\partial v}{\partial t}+\alpha v \frac{\partial v}{\partial \theta}=\alpha \frac{\partial (\pi^2 \sigma^2/2)}{\partial \theta}\\
v=\frac{\partial \Pi(\theta)}{\partial \theta}
\end{cases} \& \text{ b.c:}~~~
\begin{cases}
\sigma(\theta)|_{t=0}=\sigma_1(\theta) \\
\sigma(\theta)|_{t=A}=\sigma_2 (\theta)
\end{cases}
\end{equation}
Introducing the new function
$
f(t, \theta)=v(t, \theta)+i \pi \sigma(t, \theta)
$,
one can recast \eqref{eq17} in the form of the complex Hopf equation 
\begin{equation}\label{fe}
\frac{\partial f}{\partial t}+\alpha f \frac{\partial f}{\partial \theta}=0.
\end{equation}
\subsection{Solution to the equations of motion on the sphere}\label{sec:3.2}

The solution to  \eqref{eq17} can be found exactly and non-perturbatively in $\tau$ provided the trivial boundary conditions are imposed 
\begin{equation}\label{eq2001}
\sigma_1(\theta)=\sigma_2(\theta)=\delta(\theta).
\end{equation}
Insofar as the deformation reduces to the rescaling of the area, in the weak coupling phase $A<A_{c r}$
we take the semi-circle distribution ansatz 
\begin{equation}\label{eq21}
\sigma_*(t, \theta)=\frac{2}{\pi r^2(t)}\sqrt{r^2(t)-\theta^2}~~ ~ |\theta|<r.
\end{equation}
Plugging \eqref{eq21} into \eqref{eq17} we get the following relations
 \begin{equation}\label{eq21a}
 v( t, \theta)= \gamma(t) \theta,~~\dot{r}=\alpha \gamma(t),~~\dot{\gamma}(t)=-\alpha\Big(\gamma^2(t)+\frac{4}{r^2(t)}\Big).
 \end{equation}
 Therefore the Das-Jevicki Hamiltonian is  
 \begin{equation}\label{eq21b}
 H=\frac{\gamma^2 r^2}{8}-\frac{1}{2 r^2}.
 \end{equation}
Plugging \eqref{eq21a} and \eqref{eq21b} back in \eqref{eq17} we find the system of differential equations
\begin{equation}
\begin{cases}
\dot{r}=\frac{\gamma r}{\big[1+\tau\big(\frac{1}{12}+\frac{\gamma^2 r^2}{4}-\frac{1}{ r^2}\big)\big]^2}, \\
\\
\dot{\gamma}=-\frac{\gamma^2+\frac{4}{r^4}}{\big[1+\tau\big(\frac{1}{12}+\frac{\gamma^2 r^2}{4}-\frac{1}{ r^2}\big)\big]^2}
\end{cases}
~~~~ r(0)=r(A)=0
\end{equation}
that has two nontrivial solutions
 \begin{equation}\label{eqevol}
r(t)=2\sqrt{ \frac{t(A-t)}{A} b^{\pm}},~~~\text{where}~~~b^{\pm}(A,\tau)=\frac{1+\frac{ 2\tau}{A}(1+\frac{\tau}{12})\pm\sqrt{1+\frac{4 \tau}{A}(1+\frac{\tau}{12})}}{2(1+\frac{\tau}{12})^2}. 
 \end{equation}
They behave as follows $b^+\rightarrow1$ and $b^-\rightarrow0$ when  $\tau\rightarrow0$. Let us emphasize that $b^+$ exactly coincides with $b_{\infty}$ from \cite{tierz1} found perturbatively in $\tau$.
Therefore
the corresponding collective theory Hamiltonian evaluated on the solution reads as
\begin{equation}\label{eqdefHam}
\widetilde{ H}^{\pm}=\frac{1}{24(1+\frac{\tau}{12})}+\frac{1}{4\tau(1+\tau/12)}\Bigg(1\mp\sqrt{1+\frac{4\tau(1+\tau/12)}{A}}\Bigg). 
\end{equation} 

So far we have considered the dynamics of the Wilson loops on the cylinder of unit radius and length $A$. Now, redefining $\exp(\theta_i)\rightarrow\exp(\theta_iR)$ 
 we set the radius of the cylinder to $R$. Then \eqref{eqdefHam} takes the form of the energy of $T\bar{T}$-perturbed CFT  on the cylinder of radius $R$ in the zero momentum sector
\begin{equation}\label{eqenergy}
E^{\pm}_{}(R,\lambda)=\frac{R/24}{1-\lambda/24}+\frac{R}{2\widetilde{\lambda}}\Bigg(-1\pm\sqrt{1-\frac{4\widetilde{\lambda}}{2AR^2}}\Bigg),~~~ \widetilde{ H}^{\pm}=E^{\pm}(1,\lambda)
\end{equation} 
where $\lambda=-2\tau$ and $\widetilde{\lambda}=\lambda(1-\lambda/24)$.
This function solves the inviscid Burgers' equation (see \cite{BornInf})
\begin{equation}\label{eq021} 
\partial_{\lambda}E^{\pm}_{}(R,\lambda)=E^{\pm}_{}(R,\lambda)\partial_{R}E^{\pm}_{
}(R,\lambda)
\end{equation}
with the initial condition
\begin{equation} 
E^{+}(R,0)=-\frac{1}{2AR}+\frac{R}{24},~~~E^{-}_{}(R,\lambda\rightarrow 0)\rightarrow\infty.
\end{equation}
The appearance of the square root singularity in \eqref{eqenergy} is related to the well known shock wave solutions of the Burgers' equation, and it implies the presence of a limiting temperature. We leave the discussion of the thermodynamics of the deformed system until section \ref{sec:5}.

\begin{figure}[!tbp]
  \begin{subfigure}[b]{0.47\textwidth}
    \includegraphics[width=\textwidth]{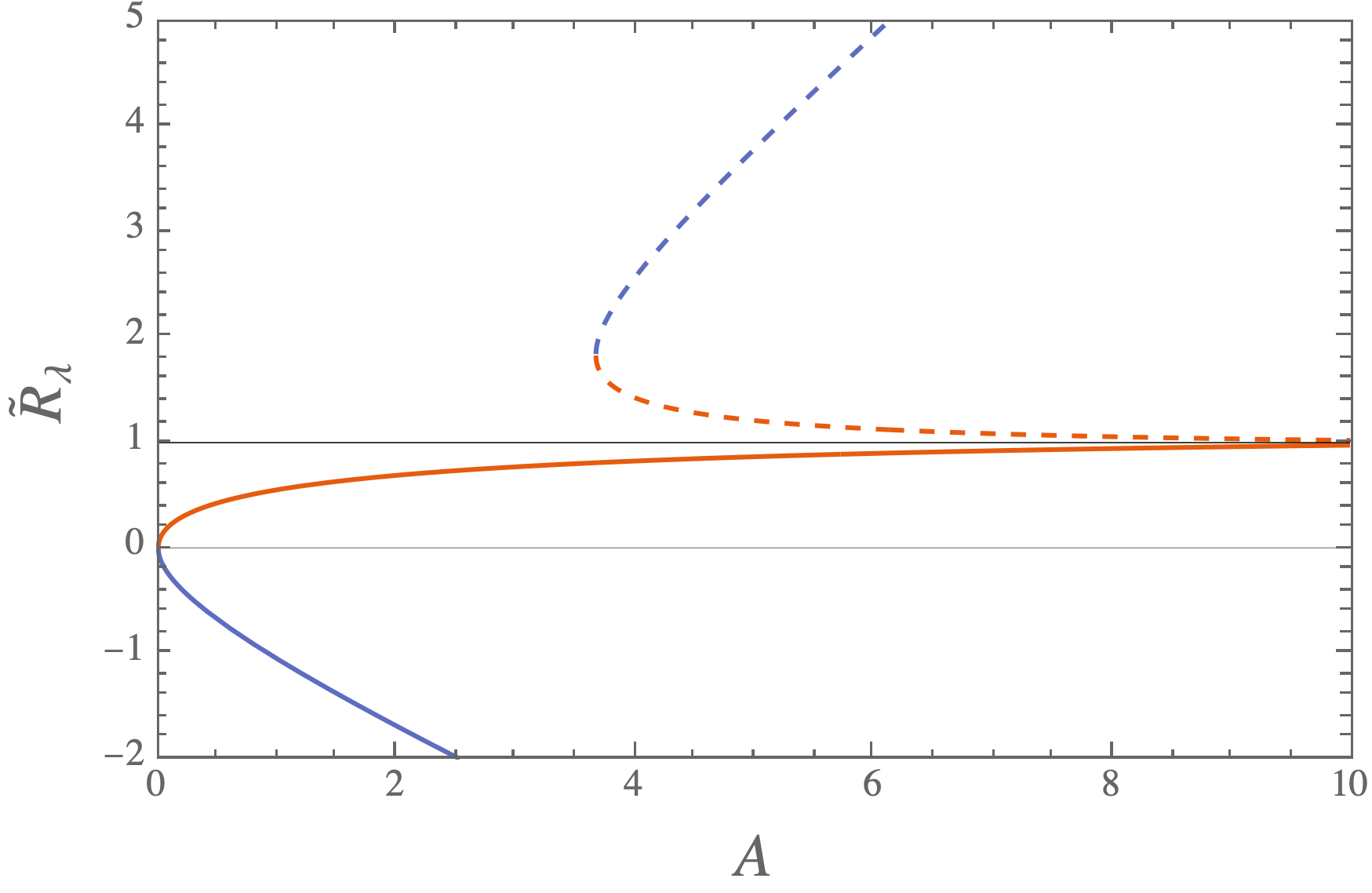}
    \caption{}
    \label{fig:f1}
  \end{subfigure}
  \hfill
  \begin{subfigure}[b]{0.46\textwidth}
    \includegraphics[width=\textwidth]{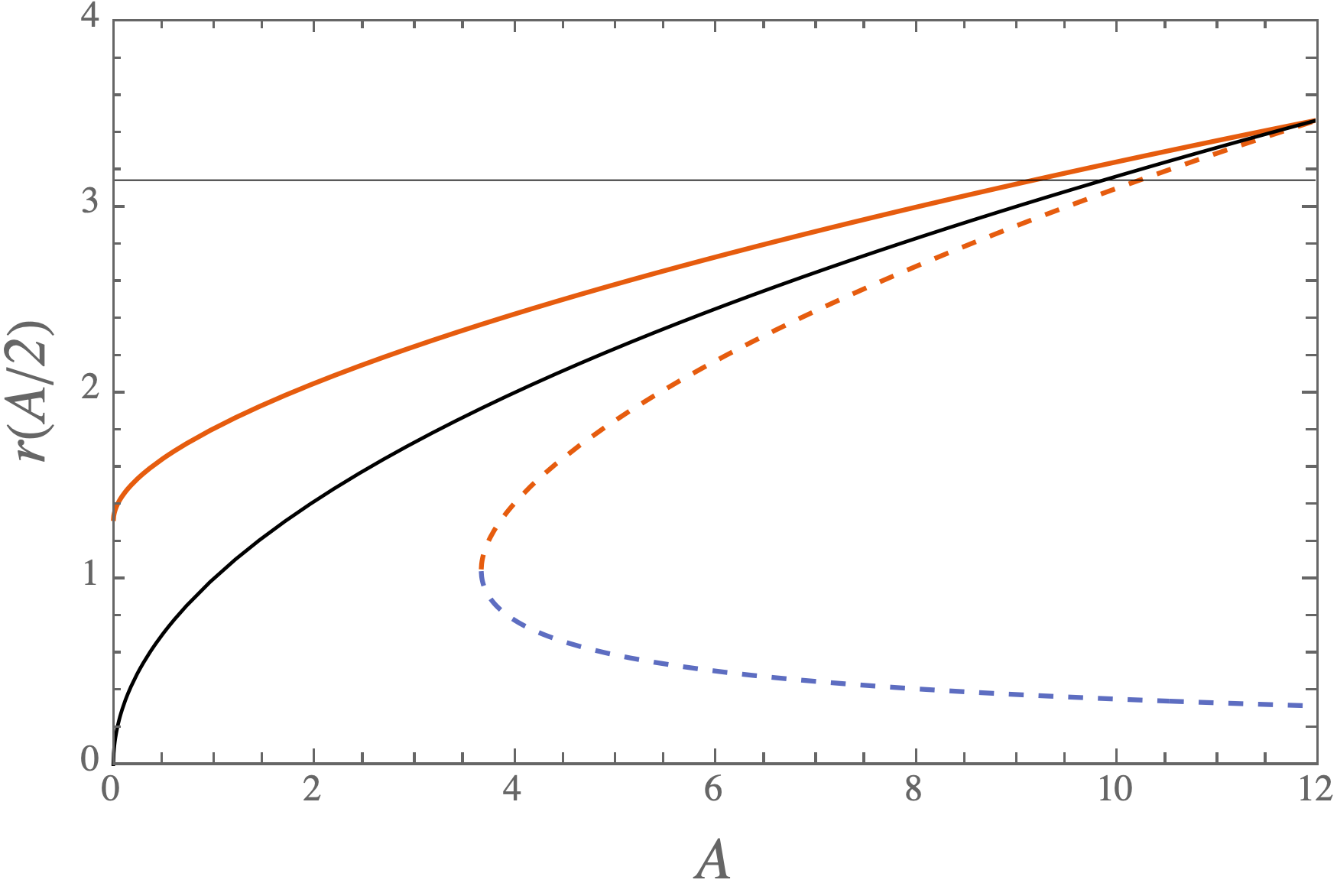}
    \caption{}
    \label{fig:f2}
  \end{subfigure}
  \caption{\label{fig:1}(a) Deformation of the radius $\widetilde{ R}^{\pm}_\lambda$, $\lambda=-2\tau$. (b) Maximum distribution $r(A/2)$ of the fluid on the unit circle. The 3-rd order phase transition takes place when $r(A/2)=\pi$.  Solid line -- $\tau>0$; dashed -- $\tau<0$. Red line -- perturbative branch; blue line -- non-perturbative branch.}
\end{figure}

$T\bar T$ deformation of non-relativistic many-body systems 
was  recently considered in \cite{cardy,jiang}. It was argued that point particles become a system of hard rods of length proportional to the deformation parameter. Indeed, using method of characteristics, one can infer the form of the perturbative branch energy
\begin{equation} 
E^{+}_{}(R,\lambda)=E^{+}_{}\big(R+\lambda E^{+}_{}(R,\lambda),0\big)
\end{equation}
and invert this equation as follows
\begin{equation} 
R^+_{\lambda}=R+\lambda E^{+}_{}(R,\lambda).
\end{equation}
Equating the radius back to $R=1$ we find that 
\begin{equation} 
{b^{\pm}}=\big(1-2\tau E^{\pm}_{}(1,-2\tau)\big)^2.
\end{equation}
The deformation comes down to the changing the radius of the circle on which the Wilson loop wraps around, while the fermionic fluid lives on the circle of radius $\widetilde{ R}^{\pm}_\lambda=1/{ R}^{\pm}_\lambda$ (see figure \ref{fig:f1}).
However this interpretation deserves for some care and need for additional arguments for its 
justification.


\section{Douglas-Kazakov phase transition}\label{sec:4}
In this section we shall generalize the 3-rd order phase transition 
in the deformed large $N$ 2d YM for the sphere \cite{tierz1} for more 
generic cases and topologies. 
Originally the transition was formulated in \cite{DK} on $S^2$ as follows. The partition function is
\begin{equation}
Z_N(A)=\sum_{R} (\dim R)^2e^{-\frac{ A}{2N}C_2(R)},~~~\dim R=\chi_R(I)=\prod_{i<j}(l_i-l_j).
\end{equation} 
In the large $N$ limit the summation over the configurations $l_i$ \eqref{eql} is replaced by a path integral
\begin{equation} 
Z(A)=\int \CMcal{D}h(x)e^{-N^2S_{eff}[\rho(h)]}
\end{equation} 
where the smooth functions are introduced
\begin{equation}
x=\frac{i}{N},~~~ h(x)=\frac{l_i}{N} ,~~~ \rho(h)=\frac{\partial x(h)}{\partial h}.  
\end{equation}
In this description, the inequalities $l_i>l_{i+1}$ transform to the restriction on the distribution $ \rho(h)$ on a Young tableau
\begin{equation}\label{eqrho}
 \rho(h)\leq 1.  
\end{equation}
A competition between the Vandermonde and Casimir terms in the effective action determines a saddle density
\begin{equation} \rho_s(A,h)=\frac{A}{2\pi}\sqrt{\frac{4}{A}-h^2}.
\end{equation}
The violation of the condition \eqref{eqrho} at $A>\pi^2$ is referred to as Douglas-Kazakov transition on $S^2$. From the perspective of the 1d fluid from section \ref{sec:3.2} the transition occurs when $\text{supp}[\sigma_*(A/2, \theta)]$ takes the whole circle $(-\pi,\pi)$ \cite{GroMat}. In the large area phase $A>\pi^2$ the eigenvalues of the Wilson loop condense around $\theta=\pm\pi$.

Let us return to the deformed YM on $S^2$. In \ref{sec:3.2} we found the semi-support
 $
r(t)=2\sqrt{ \frac{t(A-t)}{A} b^{\pm}}, 
$
that reaches its maximum at $t=A/2$. Thus the critical area equation (see figure \ref{fig:f2})
\begin{equation}
  A_{cr}b^{\pm}(A_{cr})=\pi^2 
\end{equation}
and its solution reads as
\begin{equation}
A_{cr}=\pi^2\Big[1-\tau\Big(\frac{1}{\pi^2} -\frac{1}{12}\Big)\Big]^2,~~~ \text{where} 
  ~~ \begin{sqcases}
      -\frac{12\pi^2}{12+\pi^2}<\tau<\frac{12\pi^2}{12-\pi^2} & \text{if}~~b^+ \\
     \tau< -\frac{12\pi^2}{12+\pi^2}\cup \frac{12\pi^2}{12-\pi^2}<\tau & \text{if}~~b^- 
    \end{sqcases}
\end{equation}
The change in the critical area can be clearly understood from the deformation of the radius. The circle shrinks whenever $\tau>0$, and the fluid fills all available space in less time $A$.
The large area phase $A>A_{cr}$ solution was considered in \cite{tierz1}.

\subsection{Phase transition for the disk topology}\label{sec:4.1}

Now consider $T\bar{T}$-YM on a disk with boundary holonomy $U$
\begin{equation} \label{3.8}
Z_N(U|A,\tau)=\sum_R \dim(R)\chi_R(U)e^{-\frac{ A}{2N}\frac{C_2(R)}{1-\tau C_2(R)/N^3}}.
\end{equation} 
In the large $N$ limit the eigenvalue distribution is a smooth function $\sigma(\theta)$.
To find the saddle density $\rho_s(A,h)$ we express the character through the large $N$ asymptotic of the Itzykson-Zuber integral \cite{Matytsin,Kazak}
\begin{equation} 
\chi_R(U)=I(h,i\theta)\Delta(h)\frac{\Delta(i\theta)}{\Delta(e^{i\theta})},~~~I[h,\theta]=e^{N^2F_I[\rho(h),\sigma(\theta)]}
\end{equation}
which yields for the effective action
\begin{equation} 
S_{eff}[\rho,\sigma]=-\int dh\rho(h)\int dv\rho(v)\log|h-v|-\frac{3}{2}+\frac{A}{2}\sum_{j=0}^{\infty}\tau^j\Big[\int dh\rho(h)h^2-\frac{1}{12}\Big]^{j+1}-F_I[\rho,\sigma].
\end{equation} 
Variation with respect to $h$ provides the saddle point equation
\begin{equation} \label{eq211}
\CMcal{F}(h)=Ab_{\infty}h-2\CMcal{V}(h)
\end{equation} 
where 
\begin{equation}
\CMcal{F}=\frac{\partial}{\partial h}\frac{\delta F_I[\rho,\sigma]}{\delta\rho},~~~\CMcal{V}=\fint\frac{dv\rho(v)}{h-v},~~ \text{and}~~ b_{\infty}=\sum_{j=0}^{\infty}(j+1)\tau^j\Big[\int dh\rho(h)h^2-\frac{1}{12}\Big]^{j}.
\end{equation}
Using the properties of a IZ integral Kazakov and Wynter proved in  \cite{Kazak} the relation
\begin{equation} \label{eq311}
\Theta(\theta)=-\oint\limits_{C}\frac{dh}{2\pi}\log(\theta -i\CMcal{F}(h)-iH(h)),
\end{equation} 
$H(h)=\int\frac{dv\rho(v)}{h-v}$ and $\Theta(\theta)=\int\frac{dv\sigma(v)}{\theta-v}$. 
 We plug the saddle $\CMcal{F}$ from \eqref{eq211} into \eqref{eq311} 
and obtain \begin{equation}
\Theta(\theta)=-\oint\limits_{C}\frac{dh}{2\pi}\log(\theta -iAb_{\infty}h+iH(h)).
\end{equation}
Expanding the contour around the cut one can get
\begin{equation}\label{eq37}
\Theta(\theta)=\frac{\theta}{Ab_{\infty}} -i\sum_{zeros}h(\theta)
\end{equation}
where the sum runs over all the $h(\theta)$ satisfying the condition
\begin{equation} \label{eq38}
\theta-iAb_{\infty}h(\theta)+iH(h(\theta))=0.
\end{equation} 
In the case of a non-singular $\rho(h)$ this equation has a unique solution, therefore
\begin{equation}\label{eq40}
i  Ab_{\infty}\Theta(\theta(h))=H (h).
\end{equation}

Now we can proceed according to \cite{tierz1} and look for the saddle density $\rho_s(A,h)$ perturbatively in $\tau$. Consider as an example the semicircle distribution
\begin{equation} 
\sigma(\theta)=\frac{2}{\pi c}\sqrt{c-\theta^2},~~~\Theta(\theta)=\frac{2}{c}\big(\theta-\sqrt{\theta^2-c}\big).
\end{equation}
At zero order  the eq. \eqref{eq37} takes the form
\begin{equation}
\Theta(\theta)=\frac{\theta}{A} -ih(\theta).
\end{equation}
We can invert it to get 
\begin{equation} 
\theta(h)=ih\Big(\frac{1}{A}-\frac{c_h}{4}\Big)-\frac{c_h}{2i}\sqrt{h^2-\frac{4}{c_h}},~~~c_h=\frac{A}{1-c/4}.
\end{equation}
Then, using \eqref{eq38} and the Sokhotski–Plemelj formula,
one obtains
\begin{equation} 
\rho_s^{(0)}(A,h)=\frac{c_h}{2\pi}\sqrt{\frac{4}{c_h}-h^2}.
\end{equation}
At order $\CMcal{O}(\tau^k)$ the area is renormalized as follows
\begin{equation}\label{eqp1}
A\rightarrow Ab_k,~~~b_k=\sum_{j=0}^{k}(j+1)\tau^j\Big[\int dh\rho_s^{(k-1)}(A,h)h^2-\frac{1}{12}\Big]^{j}%
\end{equation}
where the distribution $\rho_s^{(k-1)}(A,h)$ is defined at (k-1)-th step. When k goes to infinity one finds that $b_{\infty}$ is determined by the equation 
\begin{equation}
b_{\infty}=\frac{1}{\big[1-\tau\big(\frac{1}{c_hb_{\infty}} -\frac{1}{12}\big)\big]^2}
\end{equation}
that has two nontrivial solutions
 \begin{equation}
b_{\infty}^{\pm}=\frac{1+\frac{2 \tau}{c_h}(1+\frac{\tau}{12})\pm\sqrt{1+\frac{4 \tau}{c_h}(1+\frac{\tau}{12})}}{2(1+\frac{\tau}{12})^2}.
 \end{equation}

Let us get back to a general distribution $\sigma(\theta)$ and  try to get the critical area of the disk. The condition on the critical area is 
 \begin{equation}\label{eqdef}
\rho_s^{(k)}\big(A^{(k)}_{cr},0\big)=1
\end{equation} 
where $\rho_s^{(k)}$ denotes a saddle density corresponding to $\sigma(\theta)$ in the small area phase and at the k-th order of the perturbation procedure \eqref{eqp1}. The $A^{(k)}_{cr}$ is the critical area calculated at the same order.
Concretely, setting $h=0$ in eqs. \eqref{eq37} and \eqref{eq38} we get the expression
 \begin{equation} 
A^{(k)}_{c r}\sum_{j=0}^{k}(j+1)\tau^j\Big[\int dh\rho_s^{(k-1)}(A^{(k-1)}_{c r},h)h^2-\frac{1}{12}\Big]^{j}=\frac{\pi}{\Theta(\pi)}.
\end{equation} 
Now taking into account \eqref{eqdef} and \eqref{eqp1} we obtain
 \begin{equation}
\rho_s^{(k)}\big(A^{(k)}_{cr},h\big)=\rho_s^{(0)}\big(A^{(0)}_{cr},h\big) ~~~\forall~k.
\end{equation}
Therefore the final expression for the critical area of the disk is
 \begin{equation} 
A^{(\infty)}_{c r}=A^{(0)}_{c r} \bigg[1-\tau\Big(\int dh \rho_s^{(0)}\big(A^{(0)}_{cr},h\big) h^2  -\frac{1}{12}\Big)\bigg]^2,~~~A^{(0)}_{c r}=\pi\Big(\int \frac{\sigma(\theta)}{\pi-\theta} d \theta\Big)^{-1}.
\end{equation} 
As we have seen the eq. \eqref{eq37}
can be easily inverted in the case of the semicircle distribution of the eigenvalues. The main issue is finding  an undeformed saddle density $\rho_s^{(0)}(A,h)$ for general $\sigma(\theta)$. The authors of \cite{GroMat,Matytsin} reduced it to solving a system of functional equations
\begin{equation} 
\begin{cases}
G_+(G_-(x))=G_-(G_+(x))=x, \\
\text{Im}~G_+(h+i0)=\pi\rho_s^{(0)}(A,h),\\
\text{Im}~G_-(\theta+i0)=-\pi\sigma(\theta).
\end{cases}
\end{equation} 
\section{Hagedorn density of states}\label{sec:5}
Consider thermodynamic properties of $T\bar{T}$-deformed YM on $S^2$ in the small area phase. Evaluating the large $N$ asymptotic of the partition function \eqref{eqF} for the perturbative solution 
\begin{equation} 
{F}^{}=1-\frac{1}{2}\log(Ab^{+})-\frac{A}{4}\frac{\frac{1}{Ab^{+}}-\frac{1}{6}}{1-\frac{\tau}{Ab^{+}}+\frac{\tau}{12}}
\end{equation}
we obtain an effective free energy $\CMcal{F}=-N^2F$.
The partition function can be presented as the integral  over the density of states
\begin{equation} 
e^{-\CMcal{F}}=\int\rho(E)e^{-AE}dE.
\end{equation}
In what follows, we will use a somewhat sloppy notation, calling the area $ A $ as an inverse temperature.
Performing an inverse Laplace transformation we get
\begin{equation} 
\rho^{}(E)=\frac{1}{2\pi i}\int_{c-i\infty}^{c+i\infty} dA \exp[AE-\CMcal{F}]
\end{equation}
where $c$ is any positive constant lying to the right of singularities in $e^{-\CMcal{F}}$.
The exact answer is available in the undeformed theory
\begin{equation}
\rho_0(E)=\frac{e^{3/4}}{\Gamma(\frac{N^2}{2})}\Big(E+\frac{N^2}{24}\Big)^{\frac{N^2}{2}-1}.
\end{equation}
\begin{figure}[!tbp]
  \begin{subfigure}[b]{0.52\textwidth}
    \includegraphics[width=\textwidth]{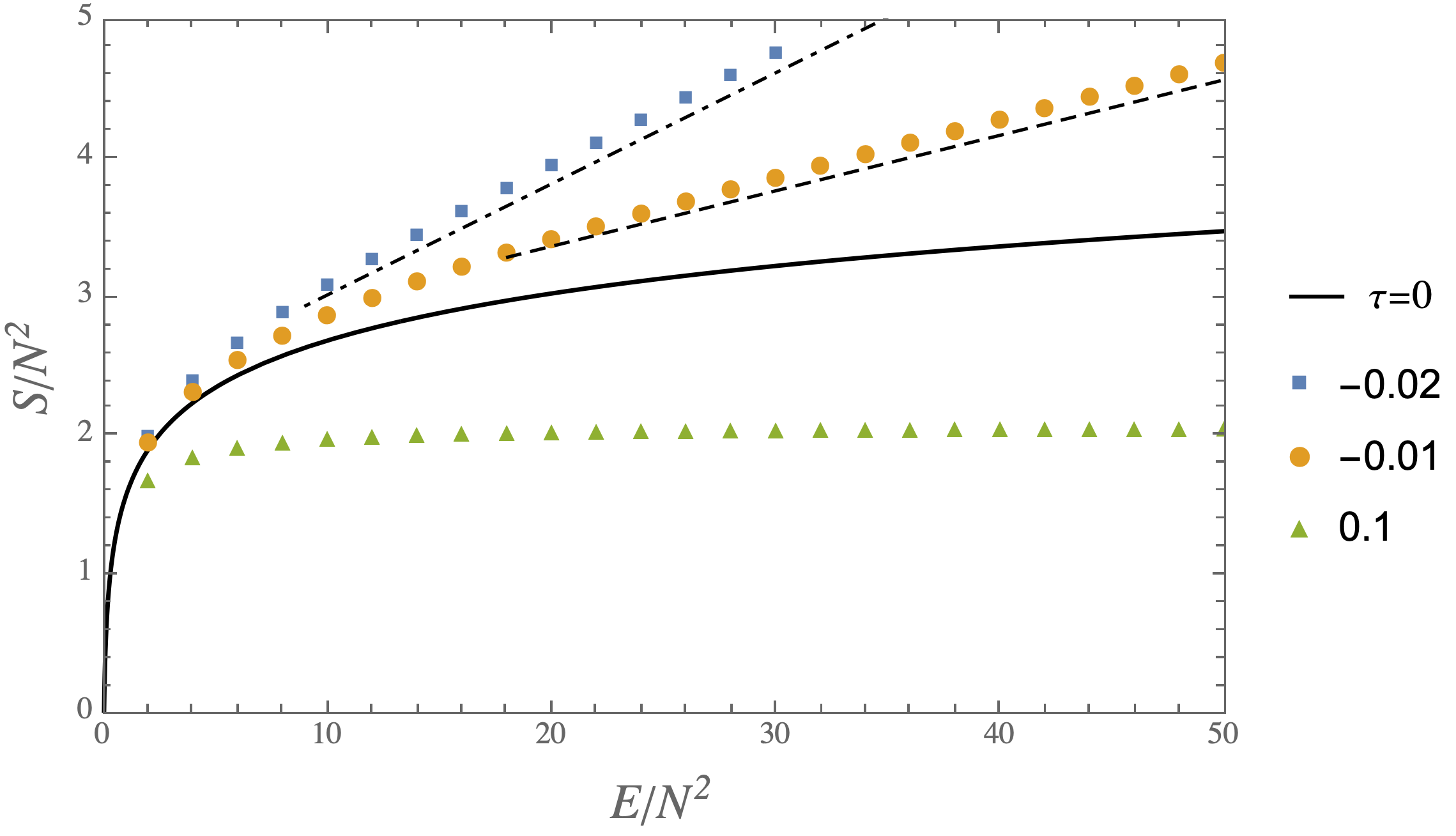}
    \caption{}
    \label{fig:f2.1}
  \end{subfigure}
  \hfill
  \begin{subfigure}[b]{0.52
  \textwidth}
    \includegraphics[width=\textwidth]{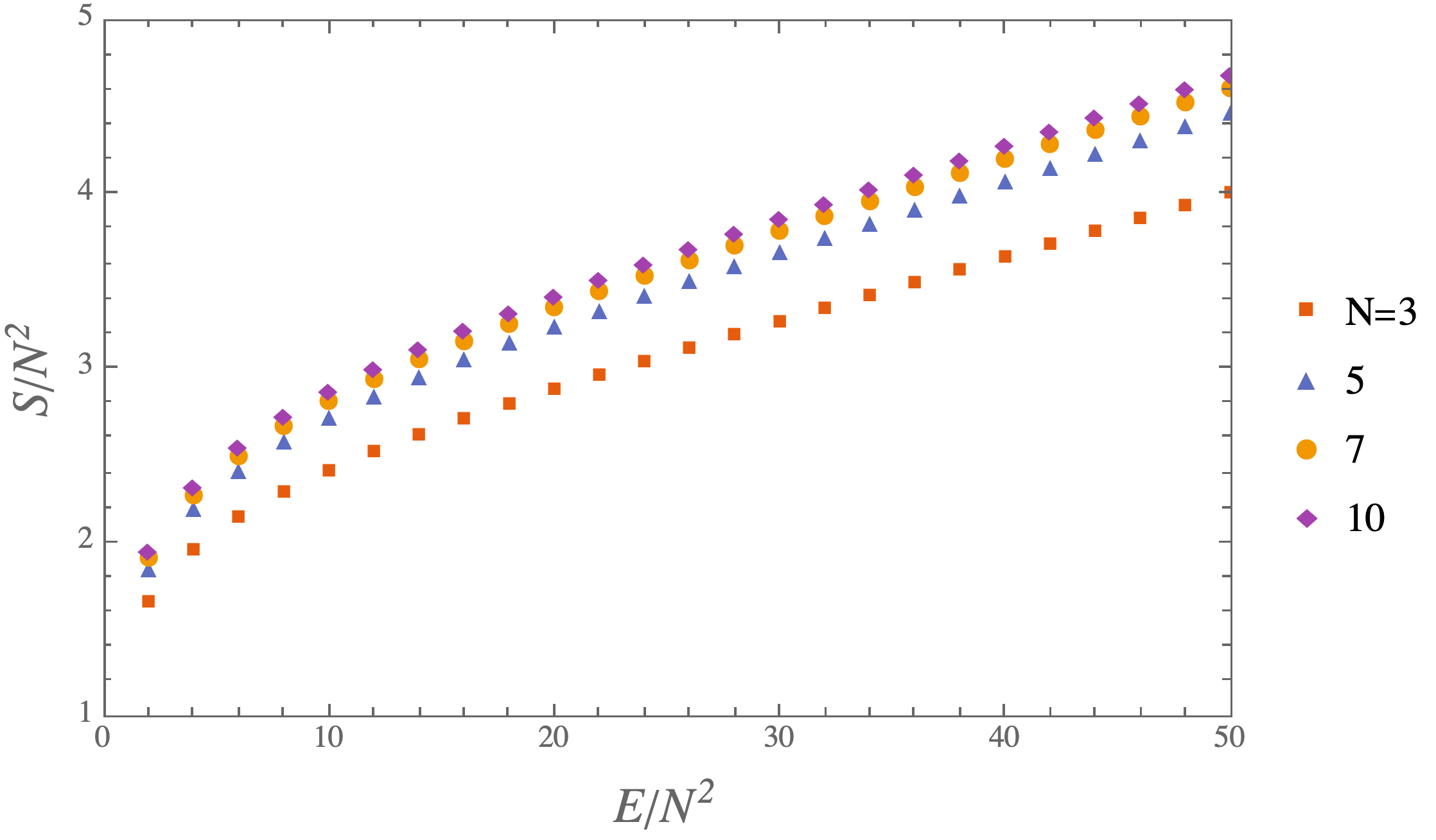}
    \caption{}
    \label{fig:f2.2}
  \end{subfigure}
  \caption{\label{fig:2}(a) Thermodynamic entropy for different values of $\tau$ at $N=10$. The slopes of the dashed and dash-dotted lines are $-4\tau(1+\tau/12)$ at $\tau=-0.01$ and $\tau=-0.02$ correspondingly. (b) The entropy at $\tau=-0.01$ for different values of $N$.}
\end{figure}
Numerical computations of the micro-canonical entropy $S=\log \rho$ for different values of the deformation parameter are shown in figure \ref{fig:2}. For negative $ \tau $ at high energies, we find the limiting temperature $ T_H $ corresponding to the root singularity in the free energy   
 \begin{equation}
T_H=\frac{1}{A_{\text{min}}}=-\frac{1}{4\tau(1+\tau/12)}.
\end{equation}
Note also that the Hagedorn temperature is higher then the temperature of the Douglas-Kazakov phase transition $T_{cr}=1/A_{cr}$ whenever $0>\tau>-\frac{12\pi^2}{12+\pi^2}$.
The significant growth of the energy level density is caused by the increase in the phase space of the fermionic fluid upon the deformation. On the other hand, the mapping of free non-relativistic fermions to a system of relativistic bosons has been discussed \cite{Mandal}. Thus, one could speculate that a bosonic Nambu-Goto long string appears under the deformation.
The emergence of the Hagedorn temperature, at which the energy pumped into the system is entirely spent on exciting new massive states, is an old result of string phenomenology.
The singularity in the partition function is due to the presence of a tachyonic mode in a string spectrum \cite{AtickWitten}.

For $\tau>0$ the phase space is reduced and the system gets maximum density. Sending the temperature to infinity we find that the entropy tends to the value 
 \begin{equation}
S/N^2=1-\frac{1}{2}\log\frac{12\tau}{12+\tau}.
\end{equation}
\section{First order phase transition}\label{sec:6}
As found in section 3 the deformed fluid can evolve along the two different trajectories  in the phase space \eqref{eqevol}. Therefore the full partition function reads as
\begin{equation} 
Z_{}=e^{-\CMcal{F}^+}+e^{-\CMcal{F}^-}
\end{equation}
where the functions $\CMcal{F}^+$ and $\CMcal{F}^-$ correspond to perturbative and non-perturbative branches of the free energy respectively
\begin{equation} 
\CMcal{F}^{\pm}=-N^2\bigg(1-\frac{1}{2}\log(Ab^{\pm})+\frac{A}{2}\frac{\big(\frac{-1}{2Ab^{\pm}}+\frac{1}{24}\big)+\frac{1}{24}}{1+2\tau\big(\frac{-1}{2Ab^{\pm}}+\frac{1}{24}\big)}\bigg).
\end{equation}
The non-perturbative part diverges provided $\tau\rightarrow0_+$. So we should neglect it when $\tau>0$. Therefore only the case $\tau<0$ is under our consideration in this section.

\begin{figure}[t]
\center{\includegraphics[width=0.614\linewidth]{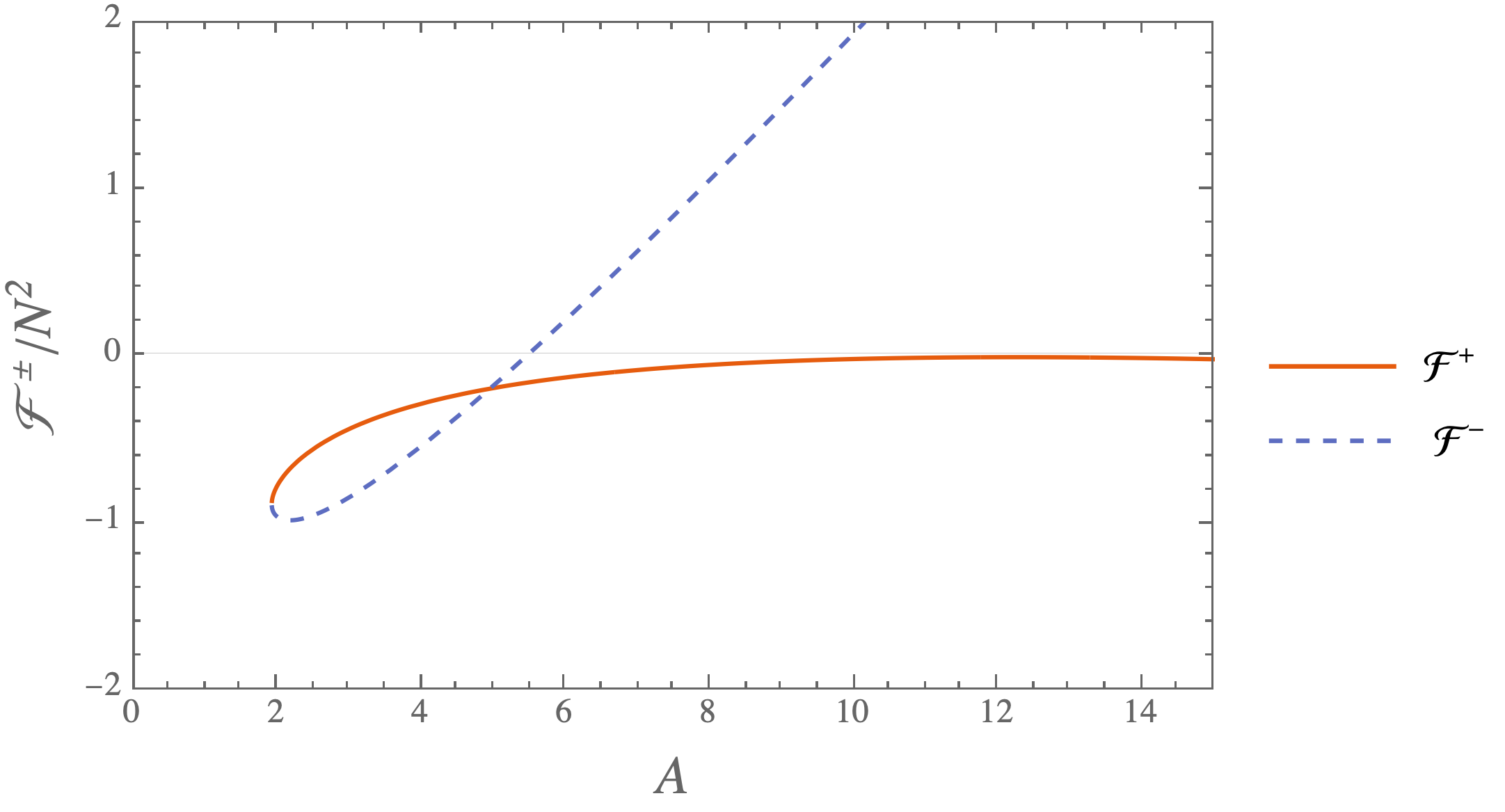}}
\caption{\label{fig:3}  Two branches for $\tau=-0.5$ in the weak coupling phase. The smallest free energy dominates in the partition function. In the large $N$ limit the first derivative with respect to $A$ of the full partition function $Z$ is discontinuous at $A_*\approx 4.948$.}
\label{ris:image2}
\end{figure}

We show that $T\bar T$-YM exhibits the phase transition between the two configurations.
As it follows from the figure \ref{fig:3} the perturbative branch dominates for the large area while the first order phase transition to the non-perturbative one occurs at $A_*$.
It is worth emphasizing that the free energy of the non-interacting hard-rod gas as a function of temperature is described by the Lambert $W$ function that has two real branches as well. However, there is no phase transition between them.
\begin{figure}[t]\label{fig:4}
\center{\includegraphics[width=0.62\linewidth]{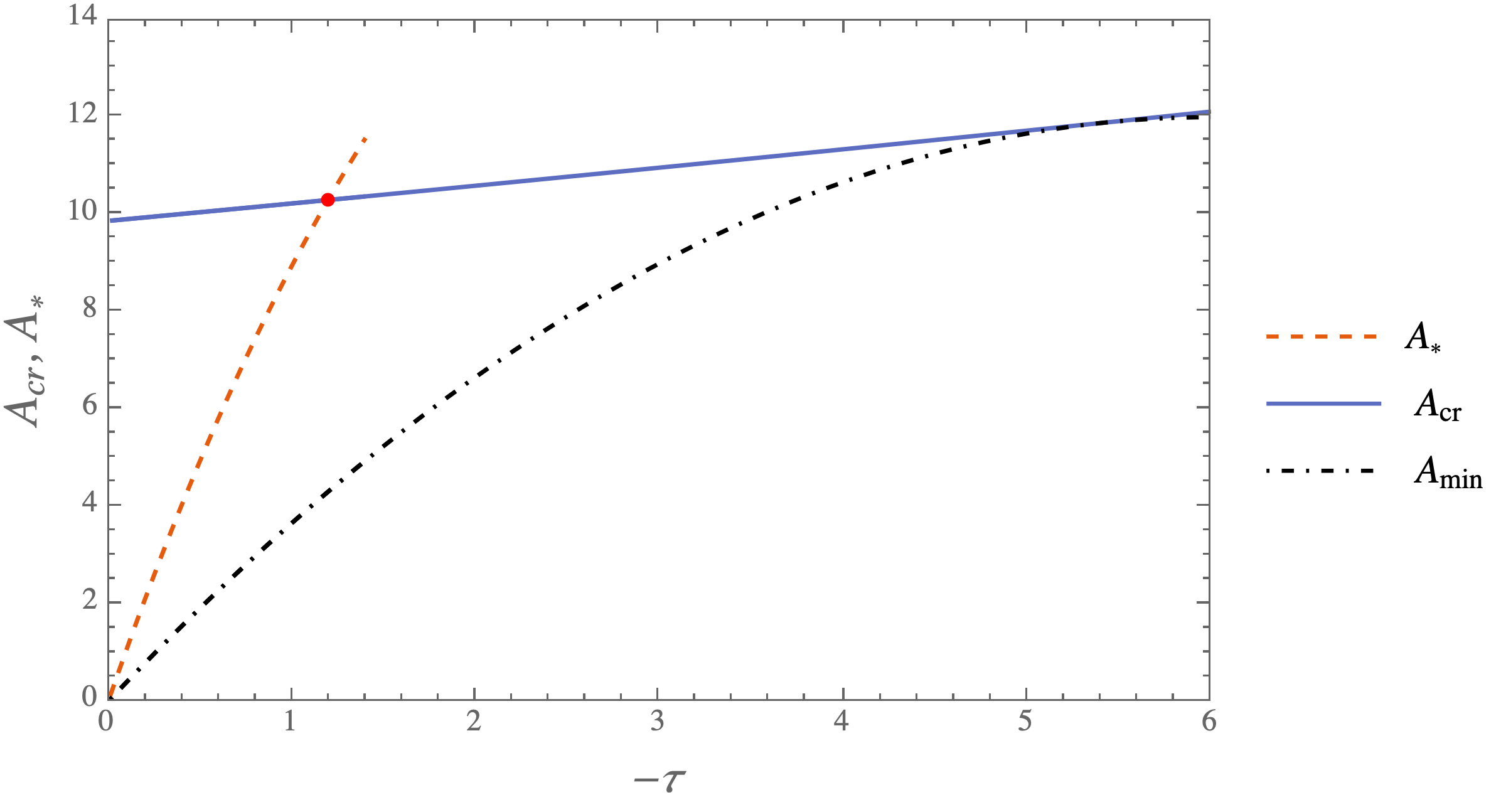}}
\caption{\label{fig:4}  The location $A_{cr}$ of the Douglas-Kazakov and $A_{*}$ of the 1-st order phase transitions as functions of $\tau$. The intersection point is  $\tau_*\approx-1.196$. $A_{\text{min}}$ corresponding to the root singularity is shown.}
\label{ris:image2}
\end{figure}

Consider the relative position of the lines of the 1-st and the 3-rd order phase transitions. The corresponding transition points $A_{cr}$ and $A_*$ are shown in figure \ref{fig:4}. We detect their coincidence at $\tau_*\approx-1.196$. If $\tau\leq\tau_*$ the non-perturbative branch completely dominates in the weak coupling phase and the full partition function does not undergo the 3-rd order transition anymore. To evaluate the point of the 1-st order transition $A_*$ slightly beyond the $A_{cr}$ we can use the weak coupling phase solution for the perturbative branch since the free energy function is quite smooth at $A_{cr}$.
\section{Comment on $T\bar{T}$-deformed q-YM}\label{sec:7}
In this section we present a few comments on the recent paper \cite{tierz2}. We analyze the peculiarities of the DK phase transition in $T\bar{T}$-deformed q-YM on a sphere for $p\leq 2$. In particular, we find that the transition does exist even for $p=1$, and the critical area turns out to be doubly degenerate. 

\begin{figure*}
        \centering
        \begin{subfigure}[b]{0.475\textwidth}
            \centering
            \includegraphics[width=\textwidth]{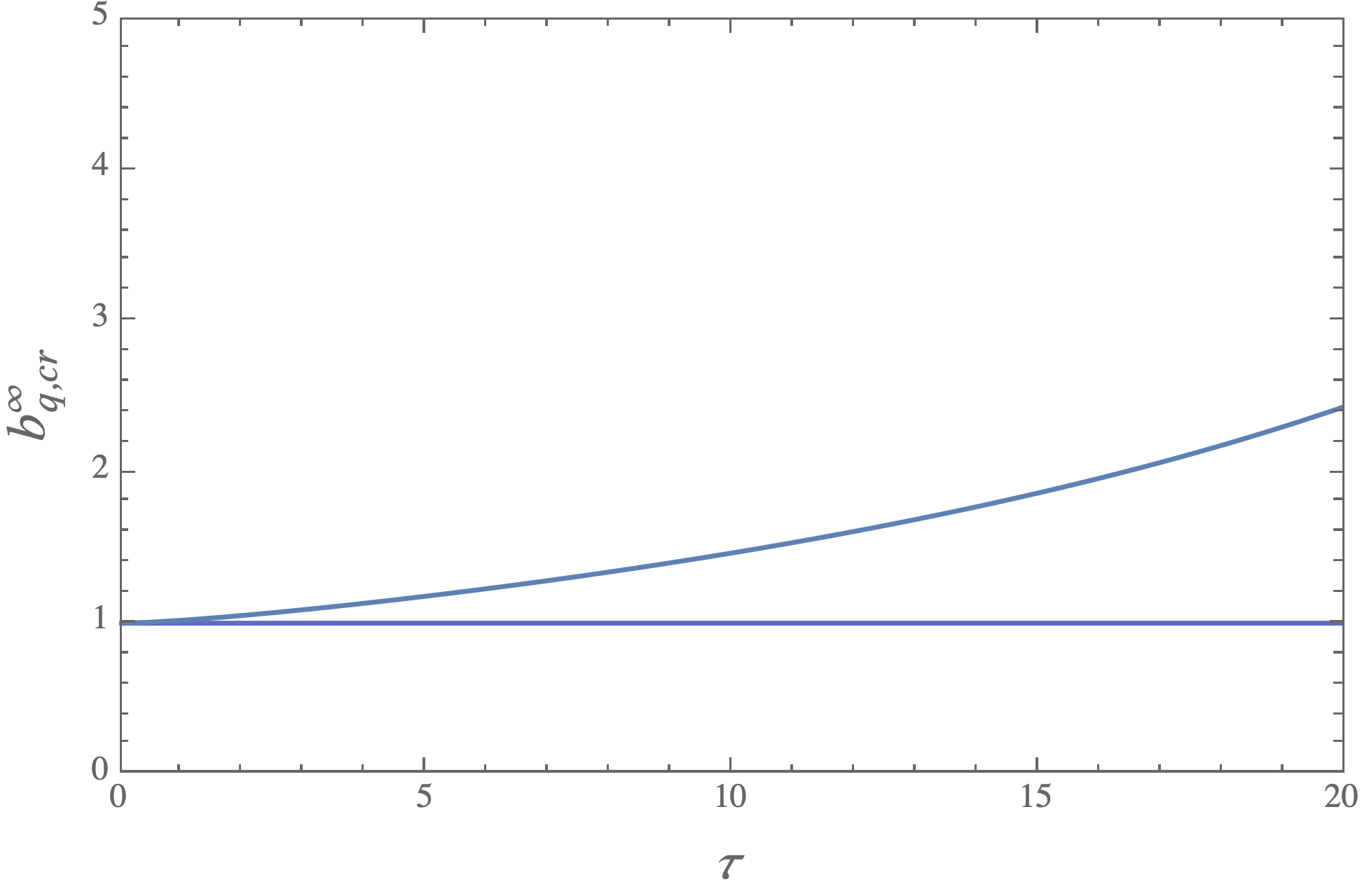}
            \caption[Network2]%
            {{\small $p=2$. There is one solution $b_{q,cr}^{(\infty)}> 1$}}    
            \label{fig:5a}
        \end{subfigure}
        \hfill
        \begin{subfigure}[b]{0.475\textwidth}  
            \centering 
            \includegraphics[width=\textwidth]{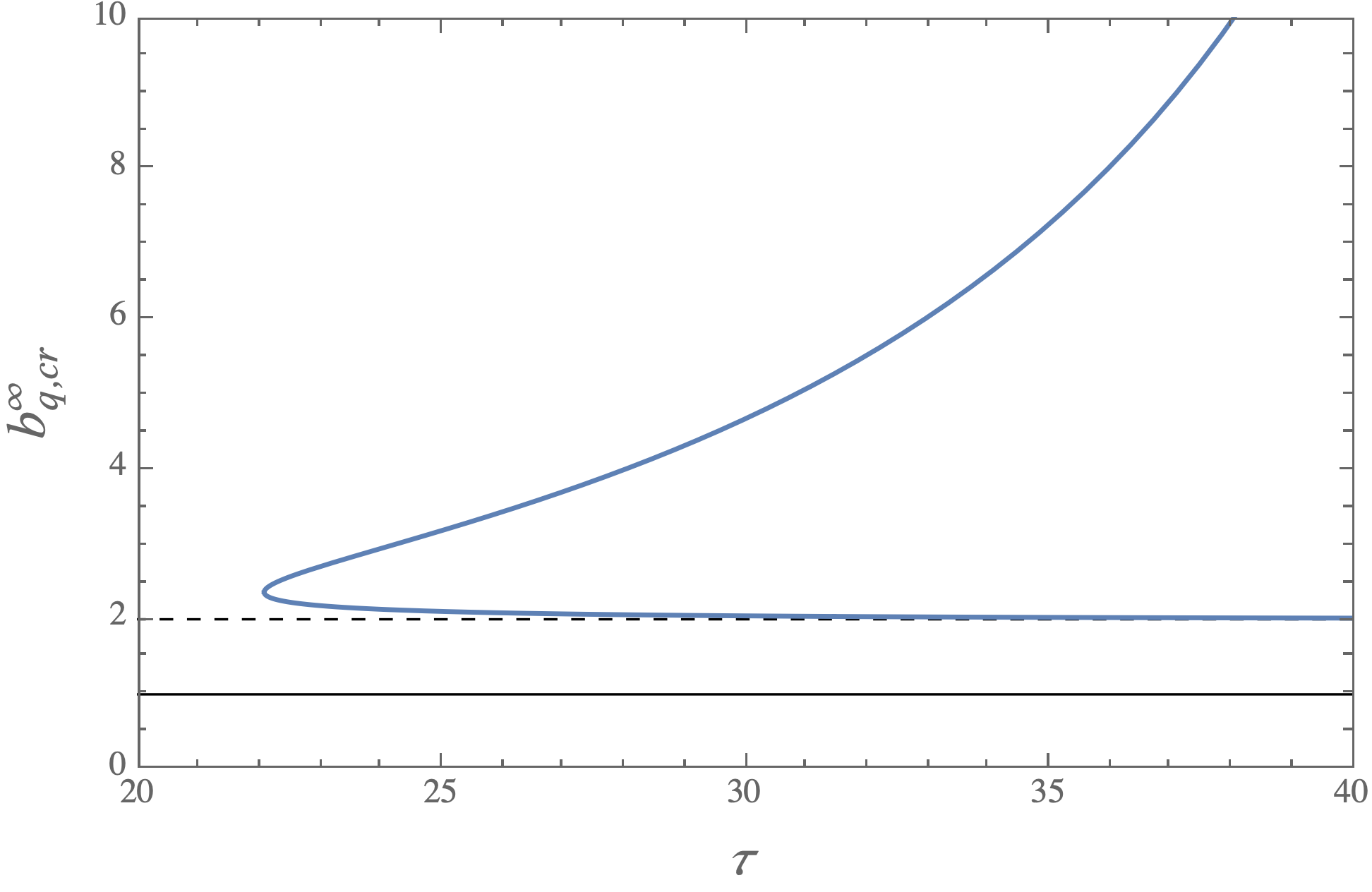}
            \caption[]%
            {{\small $p=1$. There are two solutions $b_{q,cr}^{(\infty)}> 2$}}    
           \label{fig:5b}
        \end{subfigure}
        \vskip\baselineskip
        \begin{subfigure}[b]{0.475\textwidth}   
            \centering 
            \includegraphics[width=\textwidth]{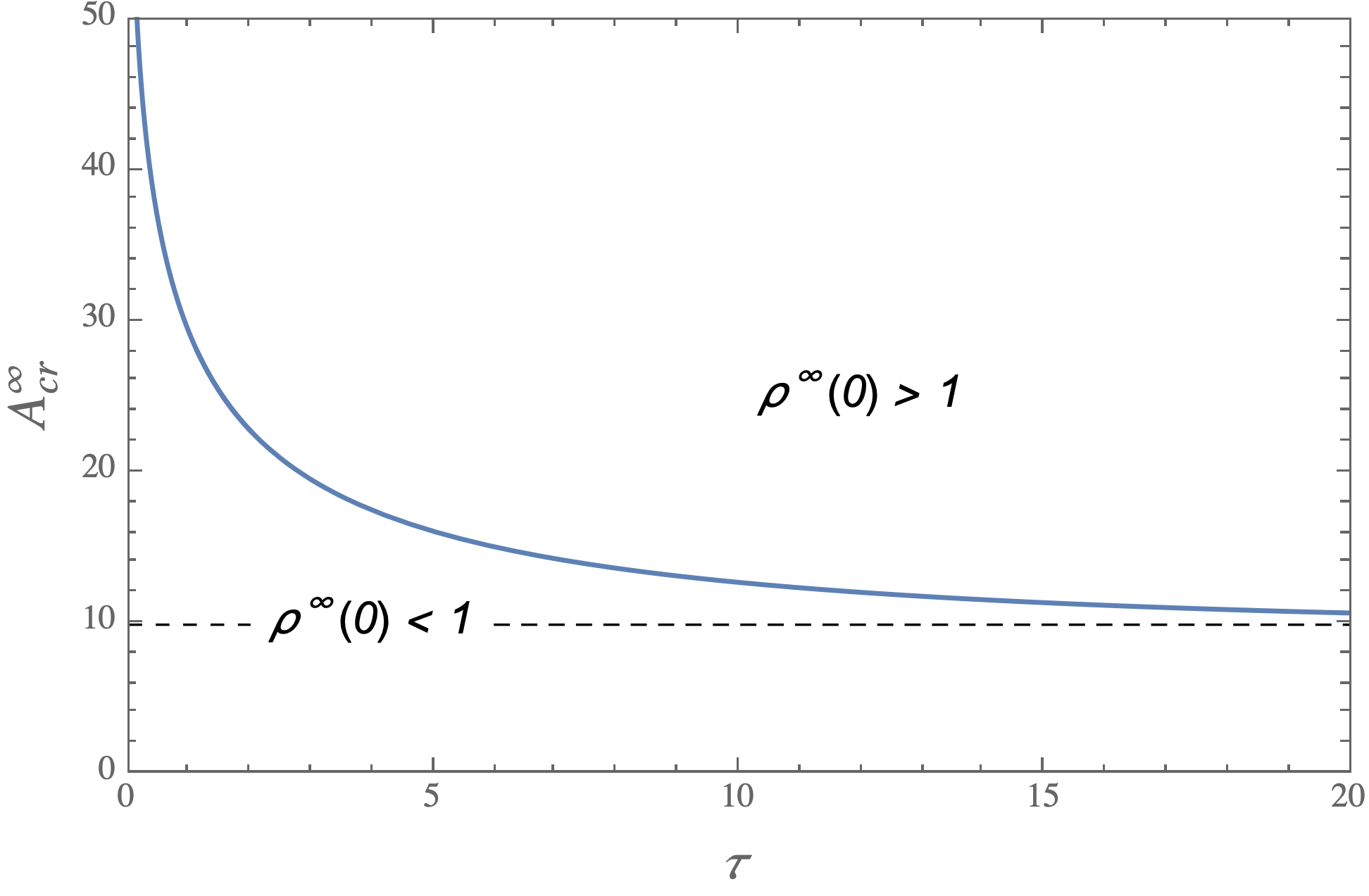}
            \caption[]%
            {{\small $p=2$. Dashed line -- $\pi^2$}}    
             \label{fig:5c}
        \end{subfigure}
        \hfill
        \begin{subfigure}[b]{0.475\textwidth}   
            \centering 
            \includegraphics[width=\textwidth]{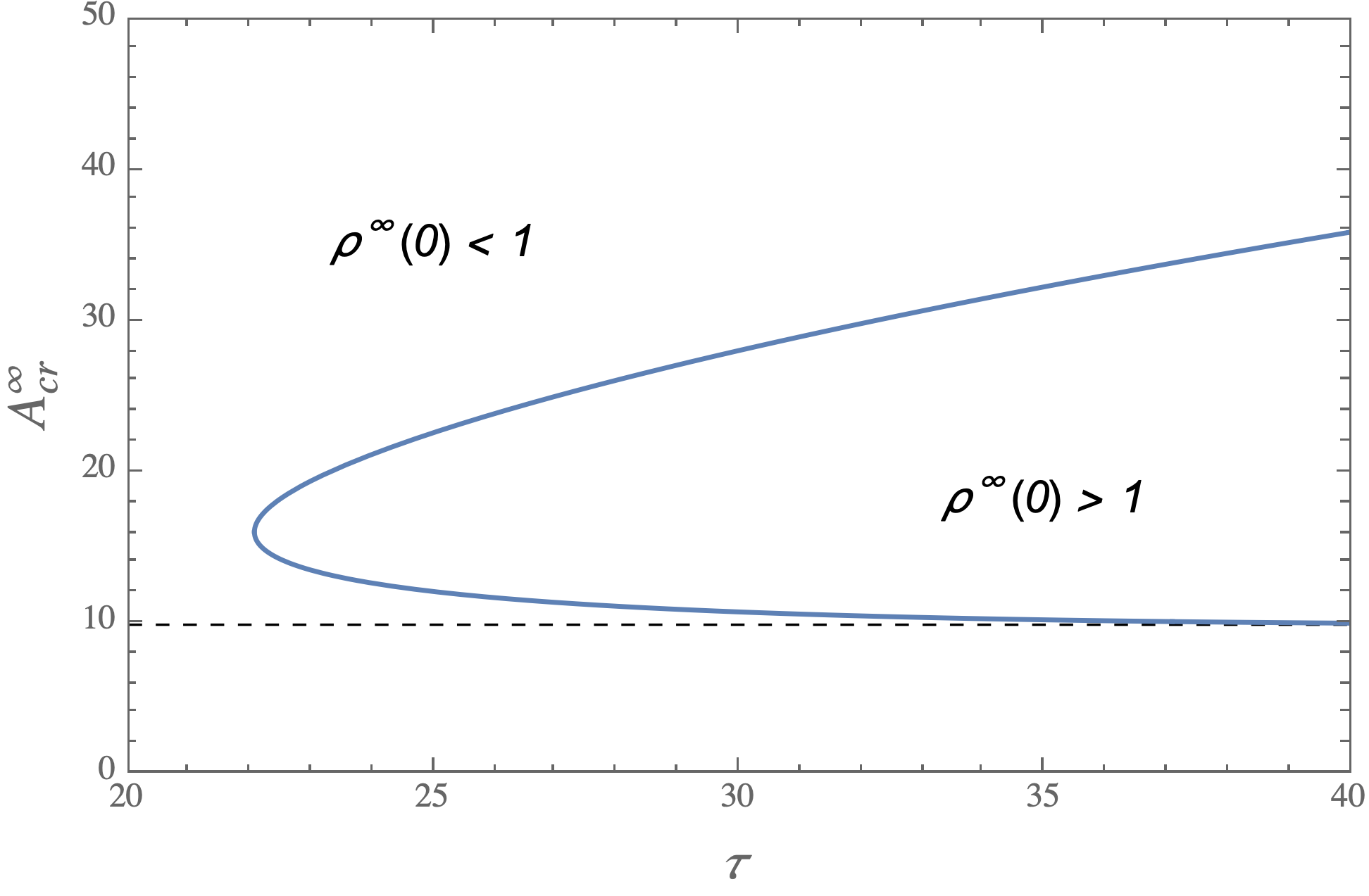}
            \caption[]%
            {{\small $p=1$. $A^{(\infty)}_{cr}$ exists for $\tau>22.06...$}}    
            \label{fig:5d}
        \end{subfigure}
        \caption[ The average and standard deviation of critical parameters ]
        {\small The solutions to the eq. \eqref{eq003} at $p=2,1$. The line $A^{(\infty)}_{cr}$ separates the one-cut phase $\rho_s^{(\infty)}\big(A,p,0\big)<1$ from the two-cut $\rho_s^{(\infty)}\big(A,p,0\big)>1$ phase.} 
       \label{fig:5}
    \end{figure*}

The quantum deformation and $T\bar{T}$
act on the YM partition function, in a sense, complementing each other. The former is obtained by replacing the dimensions of the representations with their quantum counterpart while the latter renormalized the Casimir element
\begin{equation}
Z_{q\text{-}T\bar{T}}= \sum (\text{dim}_q R)^2 q^{\frac{p}{2} \frac{C_2(R)}{1-\tau C_2(R)/N^3}}.
\end{equation}
A particular double scaling limit brings it to the $T\bar{T}$-YM.
In the continuous limit one obtains
\begin{equation}
S_{eff}[\rho]=- \int d h \int d v \rho(h) \rho(v) \log\Big[2 \sinh\frac{A(h-v)}{2p}\Big]+\frac{A}{2} \frac{\int d h \rho(h) h^2-\frac{1}{12}}{1-\tau\big(\int d h \rho(h) h^2-\frac{1}{12}\big)}
\end{equation}
\begin{equation}
+\frac{2p^2}{A^2}F_0^{CS}(\frac{A}{p})
\end{equation}
where
\begin{equation}
F_0^{CS}(t)=\frac{t^3}{12}-\frac{\pi^2t}{6}-\text{Li}_3\big(e^{-t}\big)+\zeta(3).
\end{equation}
See \cite{marino05} for details.
The saddle density at the k-th step of perturbation procedure is defined by the equation
\begin{equation}
pb_{q}^{(k)} h=\fint d v \rho^{(k)}_s(A,p,v) \coth\frac{A(h-v)}{2p}.
\end{equation}
Here we follow the notations from \cite{tierz2}
\begin{equation}
b_{q}^{(k)}(A,p)=\sum_{j=0}^{k}(j+1)\tau^j\Big[\int dh\rho_s^{(k-1)}(A,p, h)h^2-\frac{1}{12}\Big]^{j}.
\end{equation}
The solution is well known at zero order $\CMcal{O}(\tau^0)$ (see \cite{jafferis})
\begin{equation}
\rho^{(0)}_s(A,h)=\frac{p}{\pi}\tan^{-1}\sqrt{\frac{e^{A/p^2}}{\cosh^2{\frac{A}{2p}h}}-1}.
\end{equation}
The critical area, at which $\rho^{(0)}_s(A,0)=1$, is
\begin{equation}\label{eq01} 
A^{(0)}_{cr}=-p^2\log\cos^2{\frac{\pi}{p}}.
\end{equation}
It diverges at $p=2$. 
The deformed distribution is related to the undeformed one as follows
 \begin{equation}
\rho_s^{(\infty)}\big(A,p,h\big)=\rho_s^{(0)}\big(A,pb_{q}^{(\infty)},h\big),
\end{equation}
and the deformed critical area is
\begin{equation}\label{eq02}
A^{(\infty)}_{cr}=-\big(pb_{q,cr}^{(\infty)}\big)^2\log\cos^2{\frac{\pi}{pb_{q,cr}^{(\infty)}}}
\end{equation}
where we introduced the notation
\begin{equation}
b_{q,cr}^{(\infty)}=b_{q}^{(\infty)}(A^{(\infty)}_{cr},p).
\end{equation}
In \cite{tierz2}, it was emphasised that whenever $\tau>0$,  the inequality $b_{q}^{(\infty)}>1$ holds and for $p\leq 2$ there is the DK-like transition to the strong coupling phase. However, no definite answer has been given to the fate of this transition at $p=1$.
We solve numerically the equation for the critical parameters $b_{q,cr}^{(\infty)}$ at $p=1$ and $p=2$
\begin{equation}\label{eq003}
b_{q,cr}^{(\infty)}=\frac{1}{\Big[1-\tau\big(\int dh \rho_s^{(0)}\big(A^{(\infty)}_{cr},pb_{q,cr}^{(\infty)},h\big) h^2  -\frac{1}{12}\big)\Big]^2}.
\end{equation}
The second moment of the density takes the following form on the critical surface 
\begin{equation}\nonumber
\int dh\rho_s^{(0)}\big(A^{(\infty)}_{cr},pb_{q,cr}^{(\infty)},h\big) h^2=\frac{1}{\big(pb_{q,cr}^{(\infty)}\big)^2}\bigg(\frac{1}{3}+\frac{\pi^2+6 \text{Li}_2\big(\cos^2{\frac{\pi}{pb_{q,cr}^{(\infty)}}}\big)}{3\log^2\cos^2{\frac{\pi}{pb_{q,cr}^{(\infty)}}}}
\end{equation}
\begin{equation}\label{eq04}
-\frac{4 \text{Li}_3\big(\cos^2{\frac{\pi}{pb_{q,cr}^{(\infty)}}}\big)-4\zeta(3)}{\log^3\cos^2{\frac{\pi}{pb_{q,cr}^{(\infty)}}}}\bigg).
\end{equation}
 \eqref{eq04} tends to $1/\pi^2$ and $b_{q,cr}^{(\infty)}\rightarrow b_{cr}^{\pm}$ in the limit $p\rightarrow\infty$.
The solution to \eqref{eq003} is illustrated in figure \ref{fig:5}. At $p=2$ there is the trivial solution $b_{q,cr}^{(\infty)}= 1$, at which the critical area diverges. The other solution $b_{q,cr}^{(\infty)}> 1$ corresponds to the finite $A^{(\infty)}_{cr}$. However, if $p<2$, there are two relevant solutions $pb_{q,cr}^{(\infty)}> 2$. The case $p=1$ is shown in figure \ref{fig:5b}.
If we take $\tau>22.06...$ and increase the area of the sphere from zero to infinity, we find that the system undergoes the DK-like phase transition from the phase where the distribution has the one-cut form, $\rho_s^{(\infty)}\big(A,p,0\big)<1$, to the two-cut phase with $\rho_s^{(\infty)}\big(A,p,0\big)>1$ given by an elliptic curve, and then back to the one-cut phase.

\section{Discussion}\label{sec:8}
In this study we have considered several aspects of the $T\bar{T}$ deformed
2d YM theory at large $N$. The collective field theory is developed and the
generalization of Das-Jevicki Hamiltonian is found. The solution to the
equation of motion in collective theory for the sphere geometry has been elaborated.
It involves two branches and summation over  them in partition
function brings the first order phase transition at the specific value 
of deformation parameter.  

The large $N$ third order Douglas-Kazakov phase transition has been
analyzed for the disk and  the critical area 
as a function of the boundary holonomy for the arbitrary
value of the deformation parameter has been obtained. We have 
found numerically that for the sphere topology there is the intersection of the
critical lines for the 3-rd and 1-st order transitions. We are about to investigate the phase of the system near this point in the further work.
Some speculations concerning the Hagedorn-like behavior
in the deformed theory have been proposed in the present work, however this point needs for more detailed
analysis.

	There are several natural extensions of our study. First, we can
	insert the Wilson lines in time direction in different representations.
	The single Wilson line taken in a one-row representation yields the trigonometric Calogero system.
	At the large $N$ the collective field theory for the trigonometric Calogero system was 
	identified as bidirectional Benjamin-Ono equation \cite{abanov} and 
	it would be interesting to develop the deformation of the BO Hamiltonian.
	Similarly it would be instructive to find the modification of the
	critical area for the 3-rd order phase transition 
	for the $T\bar{T}$ deformed YM+Wilson line system. Another
	natural question along this line concerns the generalization of 
	the $T\bar{T}$-(YM+Wilson line) system to the $T\bar{T}$-(G/G+Wilson line). It would also
	be interesting to consider the YM-Higgs system \cite{gerasimov}
	when the ground state is closely related to the integrable Lieb-Liniger
	system. The $T\bar{T}$ deformation of the Lieb-Liniger system 
	has been discussed recently in \cite{jiang}.
	
	The 2d YM partition function is closely connected to the black hole 
	partition functions supplemented with the chemical potentials for 
	the electric and magnetic charges \cite{vafa,ooguri,baby}. Upon the $T\bar{T}$ deformation
	the simple correspondence does not work since the underlying fermionic
	picture for this correspondence gets ruined. Nevertheless we could 
	expect that some more complicated correspondence still works and we 
	hope to analyze this issue elsewhere. In particular some generalization
	of the vicious walkers picture for the deformed theory can be anticipated. 
	\\
	
\textbf{Acknowledgements:}	
We acknowledge the BASIS Foundation  grant 20-1-1-23-1  and grant RFBR-19-02-00214
for support.  
\appendix
\numberwithin{equation}{section}
\makeatletter 
\newcommand{\section@cntformat}{Appendix \thesection:\ }
\makeatother
\section{Derivation of the collective field theory}\label{Appendix1}
Following the  works \cite{GroMat,Matytsin} we pick up the ansatz  $ Z_N=e^{N^2{F}_N}$ to the deformed YM kernel at large $N$ and substitute it into the equation \eqref{eq6501}. We find that
\begin{equation}\label{A}\nonumber 
\bigg(\frac{1}{N^3}\sum_{k=1}^N \frac{\partial^2}{\partial \theta_k^2}+\frac{1}{12}\bigg) \big[\tilde{\Delta}(\theta^{})  Z_N\big]=\big[\tilde{\Delta}(\theta^{})  Z_N\big]\bigg(\frac{1}{N^2}\sum_{k=1}^N\Big(N\frac{\partial^2 {F}_N}{\partial\theta_k^{2}}\Big)+\frac{1}{N}\sum_{k=1}^N\Big(N\frac{\partial {F}_N}{\partial\theta_k^{}}\Big)^2
\end{equation}
\begin{equation}
+\frac{2}{N}\sum_{k=1}^NU_k\Big(N\frac{\partial {F}_N}{\partial\theta_k^{}}\Big)+\frac{1}{N}\sum_{k=1}^N\frac{1}{N^2\tilde{\Delta}(\theta^{})}\frac{\partial^2}{\partial\theta_k^{2}}\tilde{\Delta}(\theta^{})+\frac{1}{12}\bigg)
\end{equation}
where 
\begin{equation}
 U_k=\frac{1}{N}\frac{\partial}{\partial\theta_k^{}}\log \tilde{\Delta}(\theta^{})=\frac{1}{2N}\sum_{j\neq k}\cot\frac{\theta_k^{}-\theta_j^{}}{2}.
\end{equation}
The fourth term on the r.h.s of \eqref{A} gets reduced to the following form
\begin{equation}\nonumber
\sum_{k=1}^N\frac{1}{N^2\tilde{\Delta}(\theta^{})}\frac{\partial^2}{\partial\theta_k^{2}}\tilde{\Delta}(\theta^{})=\frac{1}{N}\sum_{k=1}^N\frac{1}{\tilde{\Delta}(\theta^{})}\frac{\partial}{\partial\theta_k^{}}\big[U_k \tilde{\Delta}(\theta^{})\big]
\end{equation}
\begin{equation}\label{B}
=\frac{1}{N}\sum_{k}^N\Big[\frac{\partial U_k}{\partial \theta_k}+NU_k^2\Big]=-\frac{1}{N^2}\sum_{k,j\neq k}\frac{1}{4\sin^2\frac{\theta_k^{}-\theta_j^{}}{2}}+\sum_{k}^NU_k^2.
\end{equation}
Therefore all  terms have the same order $\CMcal{O}(N^0)$ except the first one $\frac{\partial^2 F_N}{\partial\theta_k^{2}}$, which is of order $\CMcal{O}(N^{-1})$ and can be ignored. 
Now, using \eqref{B} and the equality
\begin{equation}
\sum_{k,m}\frac{\partial U_k}{\partial \theta_m}=\frac{1}{N}\sum_{m}\sum_{k,j\neq k}\bigg[\frac{\delta_{k,m}}{4\sin^2\frac{\theta_k-\theta_j}{2}}-\frac{\delta_{j,m}}{4\sin^2\frac{\theta_k-\theta_j}{2}}\bigg]=0
\end{equation}
we can make sure that 
\begin{equation}\nonumber
\bigg(\frac{1}{N^3}\sum_{m=1}^N \frac{\partial^2}{\partial \theta_m^2}\bigg)\bigg(\frac{1}{N^2}\sum_{k=1}^N\Big(N\frac{\partial^2 {F}_N}{\partial\theta_k^{2}}\Big)+\frac{1}{N}\sum_{k=1}^N\Big(N\frac{\partial {F}_N}{\partial\theta_k^{}}\Big)^2
\end{equation}
\begin{equation}
+\frac{2}{N}\sum_{k=1}^NU_k\Big(N\frac{\partial {F}_N}{\partial\theta_k^{}}\Big)+\frac{1}{N}\sum_{k=1}^N\frac{1}{N^2\tilde{\Delta}(\theta^{})}\frac{\partial^2}{\partial\theta_k^{2}}\tilde{\Delta}(\theta^{})+\frac{1}{12}\bigg)\sim\CMcal{O}(N^{-3}).
\end{equation}
We find that $\big[\tilde{\Delta}(\theta^{})Z_N\big]$ is an eigenfunction of the Hamiltonian up to $\CMcal{O}(N^{-3})$, and the following formula is satisfied in the large $N$ limit
\begin{equation}
\frac{1}{\tilde{\Delta}(\theta^{}) Z_N}\bigg(\frac{1}{N^3}\sum_{k=1}^N \frac{\partial^2}{\partial \theta_k^2}+\frac{1}{12}\bigg)^{j+1}\big[\tilde{\Delta}(\theta^{})\ Z_N\big]=\bigg[\frac{1}{\tilde{\Delta}(\theta^{}) Z_N}\bigg(\frac{1}{N^3}\sum_{k=1}^N \frac{\partial^2}{\partial \theta_k^2}+\frac{1}{12}\bigg)\big[\tilde{\Delta}(\theta^{}) Z_N\big]\bigg]^{j+1}.
\end{equation}
Let us introduce the function
\begin{equation}
{S}_N=-{F}_N-\frac{1}{2N^2}\sum_{j\neq k}\log \Big|\sin\frac{\theta_j-\theta_k}{2}\Big|-\frac{1}{2N^2}\sum_{j\neq k}\log \Big|\sin\frac{\phi_j-\phi_k}{2}\Big|.
\end{equation}
Therefore the eq. \eqref{eq6501} becomes
\begin{equation}
2\frac{\partial {{S}}_N}{\partial A}=\sum_{i=0}^\infty \tau^i \bigg(-\frac{1}{N}\sum_{k=1}^N\Big(N\frac{\partial {{S}}_N}{\partial\theta_k^{}}\Big)^2
+\frac{1}{4N^3}\sum_{k,j\neq k}\frac{1}{\sin^2\frac{\theta_k^{}-\theta_j^{}}{2}}-\frac{1}{12}\bigg)^{i+1}.
\end{equation}

Then we suppose that the functional $S_N$ has the regular large $N$ limit  ${S}$ and we make the replacement
\begin{equation}
\frac{1}{N}\sum_{k=1}^N\rightarrow\int_0^{2\pi}\sigma_1(\theta)d\theta,~~~
N\frac{\partial{S}_N}{\partial\theta_k^{}}\rightarrow\frac{\partial}{\partial \theta}\frac{\delta {S}}{\delta \sigma_1 (\theta)}\Bigg|_{\theta=\theta_k}.
\end{equation}
The trigonometric potential gives the interaction term
\begin{equation}
\frac{1}{N^3}\sum_{k,j\neq k}\frac{1}{4\sin^2\frac{\theta_k^{}-\theta_j^{}}{2}}=\frac{1}{N^3}\sum_{k,j\neq k}\frac{1}{(\theta_k^{}-\theta_j^{})^2}=\frac{1}{N^3}\sum_{k=1}^N\Big[\sum^N_{j=1,j\neq k}\frac{N^2\sigma_1^2(\theta_k^{})}{(j-k)^2}\Big]
\end{equation}
\begin{equation}
=\frac{1}{N}\sum_{k=1}^N\frac{\pi^2}{3}\sigma_1^2(\theta_k^{})=\frac{\pi^2}{3}\int_0^{2\pi}\sigma_1^3(\theta)d\theta.
\end{equation}
Here we just employed the fact that the region $|k-j|\ll N$ gives $\theta_k^{}-\theta_j^{}\approx (k-j)/(N\sigma_1(\theta^{}))$.
See appendix in \cite{GroMat} for details.
Finally we obtain 
\begin{equation}
\frac{\partial S}{\partial A}=-\sum_{i=0}^\infty (-2\tau)^i \bigg(\frac{1}{2}\int_0^{2\pi}\sigma_1(\theta)\Big[\Big(\frac{\partial}{\partial\theta}\frac{\delta S}{\delta\sigma_1(\theta)}\Big)^2-\frac{\pi^3}{3}\sigma_1^2(\theta)\Big]d\theta+\frac{1}{24}\bigg)^{i+1}.
\end{equation}
This equation can be solved under the assumption that the functional $S$ is an action of a dynamical system with the non-local Hamiltonian \eqref{eqHam}.
\\

\end{document}